\documentclass[11pt]{amsart}
\usepackage{geometry}                
\usepackage{graphicx}
\usepackage{amssymb}
\usepackage{epstopdf}
\newcommand{\R}{\mathbb{R}}
\newcommand{\diam}{\mbox{diam}}
\newcommand{\calP}{\mathcal{P}}
\newcommand{\one}{{\bf 1}}
\newcommand{\zero}{{\bf 0}}
\newcommand{\eps}{\varepsilon}
\newcommand{\ran}{\mbox{im}}
\newcommand{\sep}{\mbox{sep}}

\title[Persistence of spectral projections for stochastic operators]{Persistence of spectral projections for stochastic operators on large tensor products}
\author{R.S.MacKay}
\address{Mathematics Institute, University of Warwick, Coventry CV4 7AL, U.K.}
\email{R.S.MacKay@warwick.ac.uk}
\date{\today}                                           

\begin{document}
\begin{abstract}
In this paper it is proved that for families of stochastic operators on a countable tensor product, depending smoothly on parameters, any spectral projection persists smoothly, where smoothness is defined using norms based on ideas of Dobrushin.  A rigorous perturbation theory for families of stochastic operators with spectral gap is thereby created.  It is illustrated by deriving an effective slow 2-state dynamics for a 3-state probabilistic cellular automaton.  Some further potential applications are discussed.
\end{abstract}


\maketitle
\section{Introduction}
The problem of persistence of spectral projections for large tensor products is crucial in many domains.  
Perhaps the most significant domain is many-particle quantum systems.  In addition to condensed matter physics, this has taken on enhanced interest because of the problem of designing quantum registers for quantum computing.

Yet a parallel problem arises for stochastic systems with many components.  This paper specialises to Markov processes and mostly to discrete time, such as probabilistic cellular automata (PCA).  Then the transition operator $T$ acts on the space $F$ of real-valued continuous functions of the state of the whole system.  The space $F$ can be considered to be the tensor product of the spaces of real-valued continuous functions of the state of the individual units.

Even if the system is geometrically ergodic (meaning there is a unique stationary probability and it attracts every probability exponentially in an appropriate metric)
and the update of each unit is independent of the state outside a bounded neighbourhood, when one changes parameters in a reasonable way the stationary probability may move at a speed going to infinity with the number of units if distances between probability distributions are measured in any of the standard ways (e.g.~total variation, Jensen-Shannon, Hellinger, Kantorovich, Fisher information \cite{M1}, and Prokhorov \cite{M2}).
The solution proposed in \cite{M1} was to introduce a new\footnote{After publication of \cite{M1}, I found that Steif had proposed a solution 20 years earlier \cite{Sf}.  Although his metric is defined differently and in a restricted context, I was able to prove that it is in fact equal to mine in the finite case (Appendix to \cite{M2}), and with Armstrong-Goodall we have now proved they are equal under the general conditions for definition of mine \cite{AM}.} metric for probabilities on large product systems, christened ``Dobrushin metric'' as most of the ingredients were already in Dobrushin's work, but credit should also be given to Vasershtein \cite{V} (more commonly transliterated now as Wasserstein).  With respect to this metric, smooth variation of the stationary probability was proved for the class of geometrically ergodic PCA, uniformly in the size of the system \cite{M1}.  A slightly more sophisticated way of viewing this result is as persistence of the (rank-1) projection $P$ onto the space of stationary measures, for which complementary projection $Q=I-P$ sends constant functions to zero.  Thus $P$ is a spectral projection, a projection operator onto a subspace corresponding to a closed subset of the spectrum of $T$, whose complementary projection $Q=I-P$ is onto a complementary subspace corresponding to the disjoint closed complement of the spectrum.

A question that the work of \cite{M1} prompted is whether other spectral projections for stochastic operators might also persist uniformly smoothly in the size of the system.  Here suitable conditions are formulated and a proof of persistence is given.

Borel measures $p$ are dual to continuous functions $f$, in that one can take $p(f)$ to be the integral of $f$ with respect to $p$.  We think of functions as column vectors and measures as row vectors.
A transition operator $T$ acts to the right on functions and to the left on measures.  Transition operators preserve total probability, which can be written as preservation of the function $\one$, defined to take the value $1$ everywhere.

The outline of the paper is that firstly, Dobrushin metric is reviewed (sec.2).  Then the continuation problem for spectral projections of a class of stochastic operators is formulated and solved (sec.3).  An illustration of the result is given (sec.4), followed by a general development of second-order perturbation theory for stochastic operators (sec.5) and then a discussion of further potential applications, including to metastability (sec.6).  The paper ends with a short discussion (sec.7).

\section{Dobrushin metric}
Consider transition operators $T$ for functions and probability distributions on the product $X$ of a set of metric spaces $(X_s, d_s)$, for sites $s$ in a countable set $S$.  The spaces $(X_s, d_s)$ are assumed to be Polish (complete separable metric spaces) with bounded diameter, $\sup_{s\in S} \diam_s(X_s) < \infty$.  The product $X$ is endowed with product topology and with Borel measures.  The set of Borel probabilities on $X$ (measures $p$ satisfying $p(X)=1$ and $p(Y)\ge 0$ for all Borel subsets $Y$) is denoted by ${\calP}$.

For a function $f : X \to \R$, define its Lipschitz constant with respect to variations on site $s\in S$ by
$$\Delta_s(f) = \sup \frac{f(x)-f(x')}{d_s(x_s,x'_s)}$$
over $x,x' \in X$ differing at site $s$ and agreeing elsewhere.
Define the set $F$ of {\em Dobrushin smooth functions} to be those for which the semi-norm
$$|f|_F = \sum_{s\in S} \Delta_s(f)$$
is finite.
Define the space $Z$ of {\em zero-charge measures} on $X$ to be the signed Borel measures $\mu$ for which $\mu(X)=0$.
Define a norm
$$|\mu|_Z = \sup_{f\in F\setminus C} \frac{\mu(f)}{|f|_F}$$
on $Z$, where $C$ is the set of constant functions.
Then define the {\em Dobrushin distance} between any two $\rho, \sigma \in {\calP}$ by
$$ D(\rho, \sigma) = |\rho-\sigma|_Z.$$
This makes ${\calP}$ into a complete metric space, with diameter $\sup_{s\in S} \diam_s(X_s)$ \cite{M1}.

The point of this metric is that if $(I-T)$ is invertible on $Z$ then $T$ has a unique stationary probability $p \in \calP$ and it varies smoothly with respect to changes in $T$:
$$ p' = p T' (I-T)_Z^{-1},$$
where $'$ denotes derivative with respect to parameters and $(I-T)_Z$ means the restriction of $I-T$ to $Z$ (which is necessary to take its inverse) \cite{M1}.
Also there are conditions for invertibility of $I-T$ in terms of Dobrushin's dependency matrix, which are verifiable in relevant classes of system.

\section{Spectral projections}
Given a bounded transition operator $T$ on a space $F$ of continuous functions, a {\em spectral projection} for $T$ is a bounded operator $P$ on $F$ such that $P^2 = P$, $PT=TP$ and the parts of the spectrum of $T$ corresponding to the restrictions of $T$ to the image\footnote{commonly called the ``range'', but ``range'' is also used for the whole space into which $P$ maps.} of $P$ and the image of $Q=I-P$ (which are invariant under $T$) are disjoint.  Because they are closed and bounded, it follows that the distance between them is positive, called a {\em spectral gap}.

An example of a spectral projection is $P= \one p$ for the stationary probability $p$ for a geometrically ergodic operator $T$.  This is because $p \one = 1$ for any probability $p$, $T \one = \one$ by definition of a stochastic operator, $p T = p$ for a stationary measure, and geometric ergodicity implies the eigenspace for eigenvalue 1 is one-dimensional and the rest of the spectrum is in a disk of radius less than 1.

To extend the persistence theory to more general spectral projections, one needs to define norms on changes to transition operators and on tangent vectors to the manifold $M$ of projections.  

Note that the set $M$ of projections on $F$ is indeed a manifold (a variant of a Grassmann manifold).   Here is an outline proof, because ingredients are useful later.  Let $B(F)$ be the space of bounded linear operators $P$ on the space $F$ 
and define  $\Phi$ on $B(F)$ by 
\begin{equation}
\Phi(P) = P^2-P.
\end{equation}
So $M= \Phi^{-1}(0)$.  Given $P\in M$, let $R = \ran\ P$ and $K=\ker P = \ran\ Q$.  Both are closed.
Then $F = R \oplus K$.  Relative to this direct-sum decomposition, 
\begin{equation}
P = \left[\begin{array}{cc} I & 0 \\ 0 & 0 \end{array}\right].
\end{equation}
For an arbitrary operator $\pi = \left[\begin{array}{cc} \pi_1 & \pi_2 \\ \pi_3 & \pi_4 \end{array} \right]$ on $F$,
\begin{equation}
D\Phi_P (\pi) = \left[\begin{array}{cc} \pi_1 & 0 \\ 0 & -\pi_4 \end{array} \right].
\end{equation}
With respect to the same direct-sum decomposition, let $T_PM$ be\footnote{Note the re-use of the symbol $T$, which will signify tangent space as well as the transition operator.} the space of operators on $F$ of the form 
\begin{equation}
\left[\begin{array}{cc} 0 & \pi_2 \\ \pi_3 & 0 \end{array}\right],
\label{eq:TPM}
\end{equation}
and $N_PM$ be those of the form $\left[\begin{array}{cc} \pi_1 & 0 \\ 0 & \pi_4 \end{array} \right]$.
Consequently, $D\Phi_P$ maps $N_PM$ to itself by $\left[\begin{array}{cc} I & 0 \\ 0 & -I \end{array} \right]$, which has bounded inverse, namely itself.
So the implicit function theorem shows that $M$ is locally the graph of a $C^1$ function $\psi: T_PM \to N_PM$.  This completes the outline proof.
Furthermore, $D\psi = 0$, so $T_PM$ is the tangent space to $M$ at $P$ (hence the notation).
Note, however, that the manifold $M$ of projections has many components of different dimensions, corresponding to the rank of the projection.

Now the paper moves to the promised definition of norms on changes to transition operators and on tangents to $M$.

Firstly, a change $T'$ to a transition operator satisfies $T' \one = \zero$, where $\zero$ is the function taking value $0$ everywhere, so $T'$ takes any measure into $Z$.  One can quantify the size of its effect on $Z$ by the operator norm of its restriction to $Z$: 
$$|T'_Z|_Z = \sup_{\mu \in Z\setminus 0} \frac{|\mu T'|_Z}{|\mu|_Z} .$$  
One needs also, however, to measure the size of $\rho T'$ for measures $\rho$ outside $Z$, for example the Borel probabilities $\calP$, which can be non-zero even if $\mu T' = 0$ for all $\mu \in Z$.
A suitable quantification of this is 
$$L(T') = \sup_{p \in \calP} |p T'|_Z .$$  
These two quantities were used in \cite{M1} to define continuous change of a transition operator and smooth change of transition operator.
So take the norm
\begin{equation}
|T'|_* = |T'_Z|_Z + L(T').
\label{eq:norm}
\end{equation}
It is a norm because each part is non-negative, homogeneous and satisfies the triangle inequality, and $|T'|_*=0$ implies $T'=0$.  To see the latter, $|T'|_* = 0$ implies both $|T'_Z|_Z$ and $L(T')$ are zero.  $T'$ maps any measure into $Z$, so for any $\mu \in Z$ then $\mu T' = 0$ and for any $p \in \calP$ then $pT' = 0$.  Now any measure $\nu$ can be written as $kp + \mu$ for some $k \in \R$, $p \in \calP$ and $\mu \in Z$, so $\nu T'=0$, thus $T'=0$.

Secondly, the question of defining a norm on tangents to the manifold $M$ of projections is addressed.  Infinitesimal changes $P'$ to a projection $P$ do not necessarily map measures to $Z$.   They are characterised by $P'P+PP'=P'$, which can equivalently be written as $P'P=QP'$ or as $PP'=P'Q$ and hence in the form (\ref{eq:TPM}) with respect to the direct sum decomposition $F = R \oplus K$.

Nevertheless, for spectral projections one can restrict attention to the submanifolds $M_1, M_0$ of projections $P$ that fix $\one$ or send it to $\zero$, respectively, according as eigenvalue $1$ is or is not in the spectrum of the restriction of $T$ to the image of $P$.  In either case, tangents satisfy the additional condition $P'\one = \zero$.  Thus $P'$ satisfies the same conditions as a change to a transition operator above.
So, use the norm $|P'|_*$ on the tangent spaces to $M_0$ and $M_1$.
For the special case of $P' = \one p'$, where $p'$ is a change in a probability, $|P'|_*$ is precisely $|p'|_Z$.

So now the question of persistence of a spectral projection reduces to application of the implicit function theorem to the equation $PT=TP$ for $P$ in the manifold $M_0$ or $M_1$.  The tangent space $TM_i$ to $M_i$ ($i = 0,1$) at $P$ consists of the operators $P'$ such that $PP'+P'P = P'$ and $P' \one = \zero$.
Consider the functions $G_i: M_i \times N \to TM_i$ where $N$ is the space (affine with boundary) of transition operators $T$ taking non-negative functions to non-negative ones and satisfying $T \one = \one$, defined by the commutator
$$G_i(P,T) = [P,T] = PT-TP.$$ 
It is easily checked that the right hand side is in $TM_i$.
$G_i$ is jointly $C^1$ with derivatives
\begin{eqnarray}
\frac{\partial G_i}{\partial P} (P') &=& [P',T]  ,\\
\frac{\partial G_i}{\partial T} (T') &=& [P,T'] . \nonumber
\end{eqnarray}
Tangents $T'$ to $N$ are characterised by $T'\one = \zero$.
If $\frac{\partial G_i}{\partial P}: TM_i \to TN$ is invertible with bounded inverse then a solution $P$ to $G_i(P,T)=0$ has locally unique continuation $P(T)$ for nearby $T$ and depends $C^1$ on $T$, with
\begin{equation}
P' = - \left(\frac{\partial G_i}{\partial P}\right)^{-1} \frac{\partial G_i}{\partial T} T' .
\label{eq:7}
\end{equation}

The condition of bounded inverse boils down to a spectral gap.  To see this, take a direct-sum decomposition in which 
\begin{equation}
P = \left[ \begin{array}{cc} I & 0 \\ 0 & 0 \end{array} \right], \ T = \left[\begin{array}{cc} T_P & 0 \\ 0 & T_Q \end{array} \right].
\end{equation}
Then as in (\ref{eq:TPM}), $P'$ has the form
\begin{equation}
P' = \left[ \begin{array}{cc} 0 & U \\ V & 0 \end{array} \right].
\end{equation}
So 
\begin{equation}
\frac{\partial G}{\partial P}(P') = \left[ \begin{array}{cc} 0 & UT_Q-T_PU \\ VT_P - T_Q V & 0 \end{array} \right],
\end{equation}
and for (\ref{eq:7}) one wants to be able to solve $\frac{\partial G}{\partial P}(P')$ equal to an arbitrary element of $TM_i$ for $P'$.  The elements of $TM_i$ have the same zero-diagonal block form.  

Thus this reduces to solving two ``Sylvester equations'' $$AX-XB=C$$ for operators $X=U,V$ respectively, with $A,B$ equal to $T_P, T_Q$ or vice versa, and arbitrary bounded $C$.  There is a unique bounded solution $X$ to each of the Sylvester equations iff the spectra of $T_P$ and $T_Q$ are disjoint \cite{BR}.  
In the disjoint case, it
is automatic that the resulting inverse operator $C \mapsto X$ is bounded \cite{K}.

For practical purposes, one needs a bound on the inverse.
Define $$\sep(A,B) = \inf_{X\ne 0} \frac{\|AX-XB\|}{\|X\|}.$$  
This generalises the definition of the ``separation'' of two matrices $A,B$, which is usually defined using Frobenius norm \cite{Sw, Va}. 
Note that $\sep(B,A)$ is not necessarily equal to $\sep(A,B)$.
From the above discussion, $\sep(A,B) >0$ iff the spectra of $A$ and $B$ are disjoint.  

In some cases one can be explicit about its size.  For example, if the spectral radii $\rho_A, \rho_{B^{-1}}$ of $A$ and $B^{-1}$ satisfy $\rho_A \rho_{B^{-1}} < 1$ then for $\lambda>1$ there exist $C_A,C_{B^{-1}}$ (dependent on $\lambda$) such that $\|A^n\| \le C_A (\lambda\rho_A)^n$ and $\|B^{-n}\| \le C_{B^{-1}} (\lambda\rho_{B^{-1}})^n$.  The explicit (convergent) solution $$X=-\sum_{n=0}^\infty A^n C B^{-n-1}$$ 
of the Sylvester equation shows that
$$\sep(A,B) \ge \frac{1-\lambda^2 \rho_A \rho_{B^{-1}}}{C_A C_{B^{-1}} \lambda \rho_{B^{-1}}}.$$  
A method to obtain similar bounds for some more general forms of separation of the spectra is described in \cite{N}.

The important point is that for a family of examples with growing system size $N$, if the separation is bounded away from zero then the continuation above is uniform in $N$.  Furthermore, it can apply to infinite systems.


The same can be done in continuous time, with the transition operator replaced by a transition generator, but I do not spell it out here, save to mention that if the spectra of $A$ and $B$ lie in $\Re z \ge r_A$, $\Re z \le r_B$ respectively, with $r_A>r_B$, the explicit (convergent) solution $$X=\int_0^\infty e^{(-A+r)t}Ce^{(B-r)t} dt$$ for $r \in (r_B,r_A)$ provides a bound of the form
$$\sep(A,B) \ge \frac{r_A-r_B-2\eps}{c_A c_B}$$
for any $\eps>0$, with $c_A,c_B$ depending on $\eps$.

\section{An illustration}
\label{sec:ill}
As an example, consider a $3$-state PCA, with local state space $\{+,0,-\}$ at each site of a finite undirected graph with $N$ nodes and bounded degree, say by $m$.  The transition probabilities for the state at a given site are taken to be
\begin{equation}
\left[\begin{array}{ccc} 1-(1+\alpha n_- )\eps & (1+\alpha n_-)\eps & 0 \\
1/2 & 0 & 1/2 \\
0 & (1+\alpha n_+)\eps & 1- (1+\alpha n_+)\eps \end{array}\right],
\label{eq:star}
\end{equation}
in basis $(+,0,-)$, where $n_\pm$ are the numbers of neighbours in states $\pm$ respectively, $\alpha \ge 0$ and $\eps \in [0,(1+\alpha m)^{-1}]$.
For $\eps=0$ the system has eigenvalue $1$ with multiplicity $2^N$ and eigenvalue $0$ with multiplicity $3^N-2^N$.

From the above theory, the spectral projection to the subspace for eigenvalue $1$ has a continuation for $\eps$ small.  It is uniform in $N$ because the separation of the relevant operators can be bounded away from $0$ uniformly in $N$.
The dynamics on the image of the projection is slow because the spectrum moves from $\{1\}$ by at most $O(\eps)$.  It is still Markovian.  It might loosely be considered as an effective PCA with two states $\{+,-\}$ on each site but $R=\ran\ P$ is not spanned by $\delta$-functions on states, so such a description requires interpretation analogous to quasiparticles in quantum mechanics.

If $\alpha < 1/m$ and $\eps>0$, it is geometrically ergodic, because the whole system is.  This can be proved using Dobrushin's dependency matrix, as follows.  Take discrete metric on the local state spaces.  Then the update probability distributions $p(\sigma)$ for the state at a site given its current state $\sigma$ and the numbers $n_\pm$ are the rows of the  matrix (\ref{eq:star}) and so the variation distance for a change of state on the given site is
at most $(1-\eps)$ and for a change on a neighbouring site is at most $\alpha \eps$.  Thus the dependency matrix has $\ell_\infty$-norm at most $1-(1-m\alpha)\eps$.  If $\alpha < 1/m$ and $\eps>0$, this is less than $1$ and so the system is geometrically ergodic.

It is interesting to compute approximate dynamics on the image of the projection corresponding to the spectrum near $1$.  This is analogous to a second-order perturbation theory computation in quantum mechanics, e.g.~the derivation of the $t-J$ model from the Hubbard model \cite{Sp}.  Here, the $0$ state mediates interactions between the $\pm$ states.  A general treatment of second-order perturbation theory for families of stochastic operators is given in the next section.

\section{Second-order perturbation theory}

Under the spectral gap condition at $\eps=0$ and with the norm (\ref{eq:norm}), it has been proved above that for a smooth family of stochastic operators $T_\eps$ and a spectral projection $P_0$ for $T_0$, there is a range of $\eps$ for which $P_0$ continues smoothly to a spectral projection $P_\eps$ for $T_\eps$.
It is convenient to write $$P_\eps = \psi_\eps P_0 {\psi_\eps}^{-1}$$ for an $\eps$-dependent invertible bounded linear map $\psi$, with $\psi_0$ the identity.  There is a lot of freedom in the choice of $\psi$, the only constraints being that $\psi_\eps(\ran P_\eps) = \ran P_0$ and $\psi_\eps(\ker P_\eps)= \ker P_0$, but it is good also to take $\psi$ to be as smooth in $\eps$ as is $T$.  Then one can define $$S = \psi' \psi^{-1},$$ where $'$ denotes $\frac{d}{d\eps}$, and
the above conditions reduce to 
\begin{equation}
[S,P]=P',
\label{eq:SP}
\end{equation}
where $[,]$ again denotes the commutator.
A convenient solution is $$S = P'(P-Q),$$ which can be checked to satisfy the condition (\ref{eq:SP}).

Then it is desired to compute $\hat{T} = \psi^{-1}T\psi$ on $\ran\ P_0$.  This is the stochastic operator that represents the dynamics of $T_\eps$ on $\ran\ P_\eps$, using the coordinate system $\psi_\eps$.

Start from $\hat{T}_0 = T_0$.  The first derivative $\hat{T}' = \psi^{-1}(T'+[T,S])\psi$.  Using the above choice of $S$, one obtains $\hat{T}' = \psi^{-1}J\psi$, where $J = PT'P+QT'Q$.
The second derivative can be evaluated to $\hat{T}'' = \psi^{-1}(PT''P+QT''Q + [[T,P'],P'])\psi$.

Evaluating these at $\eps=0$, one obtains $\hat{T}$ to second order in $\eps$ as
\begin{equation}
\hat{T}_\eps = P_0 T_\eps P_0 + Q_0 T_\eps Q_0 + \frac{\eps^2}{2}[[T_0,P_0'],P_0'].
\label{eq:hatT}
\end{equation}
It is perhaps more useful to substitute $[T,P'] = [T',P]$, since the righthand side of this is readily computable, but the second occurrence of $P'$ above means that solution of this equation for $P'$ is unavoidable.

For the example of section~\ref{sec:ill}, there is already an effect at first order in $\eps$, so second-order perturbation theory is perhaps unnecessary, but it still serves as an illustration of the procedure.
$$T_0 = P_0 = \otimes_{s\in S} \left[\begin{array}{ccc} 1 & 0 & 0 \\ \frac12 & 0 & \frac12 \\ 0 & 0 & 1 \end{array} \right], T_0' = \otimes_{s\in S} \left[\begin{array}{ccc} -\beta_- & \beta_- & 0\\ 0 & 0 & 0 \\ 0 & \beta_+ & -\beta_+ \end{array} \right],$$
where $\beta_\pm = 1+ \alpha n_\pm$.
Thus $$P_0 T_\eps P_0 + Q_0 T_\eps Q_0 = \otimes_{s\in S} \left[\begin{array}{ccc} 1-\beta_-\eps/2 & 0 & \beta_-\eps/2 \\ \tfrac12 + \beta_+\eps/2 & (\beta_++\beta_-)\eps/2 & \tfrac12 + \beta_-\eps/2 \\ \beta_+\eps/2 & 0 & 1-\beta_+\eps/2 \end{array}\right].$$
To compute the second-order term, begin with $$[T_0',P_0] = \otimes_{s \in S} \left[\begin{array}{ccc} \frac12 \beta_- &-\beta_- &\frac12 \beta_- \\
\frac12\beta_- &-\frac12 (\beta_-+\beta_+) & \frac12 \beta_+ \\
\frac12 \beta_+ & -\beta_+ & \frac12 \beta_+ \end{array} \right].$$
Using $P_0P_0'=P_0'Q_0$, $P_0'$ has just four independent parameters on each site, and solving $[T,P']=[T',P]$ for them yields
$$P_0' =\otimes_{s\in S}  \left[ \begin{array}{ccc} \frac12 \beta_- & -\beta_- & \frac12 \beta_- \\
\frac12 \beta_+ & -\frac12 (\beta_-+\beta_+) & \frac12 \beta_- \\
\frac12 \beta_+ & -\beta_+ & \frac12 \beta_+ \end{array} \right].$$
To compute the $\eps^2$ term in (\ref{eq:hatT}), we can first subtract $[T_0',P_0]$ from $P_0'$, leaving site-wise
$$[[T_0,P_0'],P_0'] = \frac{\beta_+-\beta_-}{2} [[T_0',P_0], \left[\begin{array}{ccc} 0 & 0 & 0 \\ 1 & 0 & -1\\ 0 & 0 & 0\end{array}\right] ]
= \frac{\beta_+-\beta_-}{2} \left[\begin{array}{ccc} -\beta_- & 0 & \beta_-\\-\beta_- & \beta_- - \beta_+ & \beta_+ \\ -\beta_+ & 0 & \beta_+ \end{array}\right].$$

Thus the final result to second order, using the $(+,-)$ part of the $\psi_\eps$ basis, is the effective PCA
$$\hat{T}_\eps = \otimes_{s\in S} \left[\begin{array}{cc} 1-\eps\beta_-/2+\eps^2\beta_-(\beta_- -\beta_+)/4 & \eps\beta_-/2-\eps^2\beta_-(\beta_- -\beta_+)/4 \\
\eps \beta_+/2-\eps^2\beta_+(\beta_+ -\beta_-)/4 & 1-\eps \beta_+/2 + \eps^2\beta_+(\beta_+ -\beta_-)/4 \end{array} \right].$$

The outcome is a two-state model in which to leading order there is a small probability $\eps \beta_\mp/2 = \eps (1+\alpha n_\mp)/2$ per timestep for transition from state $+$ to $-$, respectively $-$ to $+$.

Perhaps a reader can suggest (and apply the method to) a more significant example.


\section{Further potential applications}
One area to which the above results might be usefully applied is metastability. 
Metastability is the phenomenon that an ergodic process may spend long times exploring restricted subsets of the support of the probability distribution, switching between them infrequently but sufficiently to achieve ergodicity in the long run.  One reference is \cite{BdH}.  

As an application of the result, one can start from the paper \cite{D} in which the hypothesis is a reversible Markov process in continuous time with spectrum in 
$[-\eps,0] \cup [-\infty,-1]$ such that each function in the image of the spectral projection $P$ for $[-\eps,0]$ is bounded, and it is proved that there is a partition of the state spaces into ``metastable regions''.  It is not clear to me, however, whether there are relevant examples satisfying the hypotheses.  One might think that Glauber dynamics of the 2D Ising model below the critical temperature would qualify, but I'm not aware that it is proved to have a spectral gap.  Nevertheless, if there are examples on product spaces then the result of the present paper proves robustness of spectral gap and hence of the phenomenon of metastability.

One could envisage the result also being useful to treat perturbation of Markov dynamics with more than one stationary distribution, for example with more than one communicating component.  Addition of some interaction between the communicating components typically makes the system have a single communicating component but the result of this paper shows there is a continuation of the spectral projection to a spectral projection with spectrum contained near $1$ and the same rank (equal to the original number of communicating components), and it gives strong control over the resulting continuation.  
More substantially, one would like to deduce something about metastability in ergodic finite versions of infinite PCA with non-unique stationary distribution.

Another possible application is to perturbations of product systems in which the units all have simple eigenvalue +1 and isolated spectrum near some $\lambda$ in the open unit disk.  The conclusion is that there persists an invariant subspace with decay constant near $\lambda$.

\section{Discussion}

It has been proved here that every spectral projection of a stochastic operator on a product space persists $C^r$-smoothly with respect to the norm (\ref{eq:norm}) for $C^r$-smooth changes in the transition operator, again measured using (\ref{eq:norm}).
This generalises the case of the rank-one projection onto a stationary distribution, treated in \cite{M1}.

Second-order perturbation theory has been developed for families of such operators and an example treated.
Potential applications have also been suggested to robustness of metastability and some other uses for multi-component stochastic processes.


\section*{Acknowledgements}
I am grateful to Nick Higham for pointers to the literature on Sylvester equations.

\end{document}